\documentclass[aps,prl,showpacs,nofootinbib]{revtex4}
 \usepackage{graphics}
 \usepackage{amsmath}
 \usepackage{epsfig}
 \usepackage[english]{babel}

\begin{document}

\title{Manifestation of the cyclo-toroid nuclear moment in anomalous conversion and Lamb shift\footnote{Submitted to Phys. Rev. Lett. June 17, 2005}}

\author{E.V.~Tkalya}

\email{tkalya@srd.sinp.msu.ru}

\address{Institute of Nuclear Physics, Moscow State University,
Ru-199999, Moscow, Vorob'evy Gory}

\date{\today}

\begin{abstract}
We offer the hypothesis that atomic nuclei, nucleons, and atoms
possess a new type of electromagnetic moment, that we call a
``cyclo-toroid moment''. In nuclei, this moment arises when the
toroid dipole (anapole) moments are arrayed in the form of a ring,
or, equivalently, when the magnetic moments of the nucleons are
arranged in the form of rings which, in turn, constitute the surface
of a torus. We establish theoretically that the cyclo-toroid moment
plays a role in the processes of the atomic shell--nucleus
interaction. The existence of this moment would explain known
anomalies in the internal conversion coefficients for $M1$
transitions in nuclei. We show also that the static cyclo-toroid
nuclear moment interacts locally inside the nucleus with the vortex
part of the atomic electron currents and this leads to an energy
shift in atomic $s_{1/2}$ states. For the hydrogen atom the value of
this shift may be comparable in order of magnitude to the present
accuracy of measurements of the Lamb shift for the $1s_{1/2}$ level.
\end{abstract}


\maketitle

\section{1. Introduction}

    The existence of anapole moment was theoretically predicted by
Zel'dovich fifty years ago \cite{Zeldovich-57}. The anapole moment
arises in the system where currents or magnetic moments of particles
create a ring-like closed distribution of magnetic field lines. A
classic example of such distribution can be a magnetic field inside
the toroidal solenoid over the surface of which poloidal currents
flow. A less trivial example is an atomic nucleus in which spins of
nucleons compose a ring with forming the corresponding configuration
of the magnetic field.  The great interest to the anapole is related
to fundamental investigations on the parity nonconservation in
atomic-nuclear interactions \cite{Bouchiat-97,Flambaum-80}. In 1997
the anapole moment of the nucleus $^{133}$Cs was discovered just in
the experiment on parity nonconservation in the atomic transition
$6s\rightarrow7s$ of Cs \cite{Wood-97}.

    Now the anapole is regarded as a dipole moment of the fundamental
family of toroid moments \cite{Dubovik-74,Dubovik-90} which naturally
arise when currents on a torus surface are described. Toroid moments
are now used in problems of current parametrization and of radiation
theory \cite{Dubovik-74}, in investigations of topological effects in
quantum mechanics (the Aharonov-Bohm effect \cite{Afanasiev-96}), in
elementary particle physics (toroid dipole momentum of neutrino
\cite{Bukina-98} and $NN$-scattering \cite{Haxton-01} for example),
in the theory of anomalous internal conversion \cite{List-81}, in
atomic physics (anapole moments of atoms \cite{Lewis-93}, atomic
emission in a condensed medium \cite{Tkalya-02}), in the physics of
nanostructures (carbon toroids \cite{Ceulemans-98}), in the physics
of spin-ordered crystals \cite{Schmid-01} etc. Thus the physics of
toroid moments is a wide area of scientific activity ranging from
elementary particles to condensed matter. At the same time, the
nucleus together with the atomic shell remain one of the most
suitable quantum object for probing the toroid moments. The reason is
as follows. The electronic wave functions for $s_{1/2}$ and $p_{1/2}$
states have large amplitudes at the origin. As a consequence, the
electronic current at such states effectively penetrates the nucleus
and interacts with the toroid moment, whose magnetic field does not
exceed the boundary of the nucleus. (This interaction mechanism is a
direct analogy of a thought experiment offered in
Ref.~\cite{Zeldovich-57} for a toroidal solenoid immersed into the
electrolyte.) Today the situation is. The static toroid dipole moment
has discovered \cite{Wood-97}. As regards the toroid dipole moment of
transition, there are evidences \cite{List-81} that this moment
results in known anomalies in the internal conversion coefficients
for $E1$ transitions in nuclei
\cite{Church-56,Green-58,Nilsson-58,Voikhanskii-59} and in the
process of nuclear excitation by electron transition in the atomic
shell \cite{Tkalya-94}. Thus, more detailed and precise
``conversion'' experiments are needed to detect the toroid dipole
moment of transition.

    In the present work we consider, on the example of the
nucleus--atomic shell system, two new effects dealt with the toroid
moments. First of all, we prove that the anomalies mentioned above
for $M1$ transitions result from a ``cyclo-toroid moment''. This
moment arises when the toroid dipole moments are arranged in a ring.
Secondary, we demonstrate a possibility of experimental investigation
of the existence of the cyclo-toroid moment by measuring of the
energy shift in atomic $s_{1/2}$ levels. The results obtained here
would be useful for more fundamental understanding of the nontrivial
distribution of the currents inside nucleus and nucleons, and for the
processes of the interaction between nucleus and atomic shell.

\section{2. Cyclo-toroid nuclear moment of transition}

    Anomalous internal conversion of $\gamma$-rays in deformed nuclei
arises because of the penetration of the electron current $j_e(r_e)$
into the nucleus ``beneath'' the nuclear current $j_N(r_N)$, i.e.
when the current coordinates satisfy the condition $r_e<r_N$. The
reason which changes the probability of the mentioned atomic-nuclear
processes originates from to the additional selection rules
\cite{Nilsson-58,Voikhanskii-59}. Nuclear and atomic matrix elements
which determine the process of de-excitation (excitation) of deformed
nuclei through the atomic shell for $r_e<r_N$ are different from
those for $r_e>r_N$. In some cases the additional selection rules
allow the nuclear transition for $r_e<r_N$ and forbid a
$\gamma$-radiative transition \cite{Voikhanskii-59} with nuclear
matrix element $\langle \hat{\cal{M}}_{LM}^{e/m}\rangle = \int d^3r_N
\, {\bf{j}}_N\cdot {{\bf{A}}^{e/m^*}_{LM}}$ entering the probability
formula for $r_e>r_N$. (Here and further ${\bf{A}}^{e/m}_{LM}$ are
the standard electric/magnetic multipoles according to the gauge of
$\text{div}{\bf{A}}^{e/m}_{LM}=0$ (see in Ref.~\cite{Eisenberg-70}),
and $j_{e,N}$ are current densities. We call $j_{e,N}$ ``currents''
for short.)

    At the beginning of 80s it was found out that the real cause of
the mentioned anomaly in the processes of an electric-dipole ($E1$)
interaction of the nucleus with the atomic shell results from the
dipole toroid moment of the transition \cite{List-81}. Toroid moments
are the components of electric moments \cite{Dubovik-74,Dubovik-90}.
Moreover, the anomaly also exists in conversion magnetic-dipole
($M1$) transitions \cite{Church-56,Nilsson-58}. Let us consider the
nature of such $M1$ anomaly.

    The nucleus matrix element corresponding to the anomalous part of
the magnetic-dipole process of the nuclear interaction with the
atomic shell is as follows \cite{Church-56,Voikhanskii-59,Tkalya-94}
\begin{equation}
\langle \hat{\cal{M}}_{1M}^{{m}^{an}} \rangle = \int d^3r
\left(\frac{r}{r_0}\right)^2 {\bf{j}}_N(r)\cdot
{{\bf{A}}^{m^*}_{1M}}(\omega,r),
\label{eq:N-an}
\end{equation}
where ${\bf{A}}^{m}_{LM}(\omega,r) =
j_L(\omega{r}){\bf{Y}}^L_{LM}({\bf{n}})$ \cite{Eisenberg-70},
$\omega$ is the energy of a nuclear transition, ${\bf{n}} =
{\bf{r}}/r$, $j_L(\omega{r})$ are the spherical Bessel functions,
${\bf{Y}}^L_{JM}({\bf{n}})$ are the vector spherical harmonics from
Ref.~\cite{Varshalovich-88}, $r_0$ is the nucleus radius. We use in
this paper the system of units $\hbar = c=1$.

    The operator structure of $\langle\hat{\cal{M}}_{1M}^{{m}^{an}}\rangle$
is well known. If the nuclear transition current in
Eq~(\ref{eq:N-an}) is the sum of convection and spin terms
${\bf{j}}_N = {\bf{j}}_N^c + {\bf{j}}_N^s$, where ${\bf{j}}_N^c =
-ie_N/2m_N (\varphi^*_f{\boldsymbol{\nabla}}\varphi_i -
({\boldsymbol{\nabla}}\varphi^*_f)\varphi_i)$ and ${\bf{j}}_N^s =
\text{curl}(\varphi^*_f {\hat{\boldsymbol{\mu}}}\varphi_i)$ (we
consider one-nucleon current for simplicity) then it is easy to get
the following expression for the anomalous nuclear matrix element
\cite{Church-56,Nilsson-58,Voikhanskii-59}
\begin{eqnarray}
\langle \hat{\cal{M}}_{1M}^{{m}^{an}} \rangle &=&
  \frac{i\sqrt{2}}{r_0^2} \int
d^3r \big({\boldsymbol\nabla}j_1(\omega{r})Y_{1M}^*({{\bf{n}}})\big)
\nonumber \\
&&
 \cdot\varphi_f^*(r)
\left(\frac{e_N}{2m_N}r^2{\hat{\bf{L}}}-
\left[{\bf{r}}({\hat{\boldsymbol{\mu}}}\cdot{\bf{r}})-
2{\hat{\boldsymbol{\mu}}}r^2 \right] \right) \varphi_i(r).
\label{eq:Operat-an}
\end{eqnarray}
In formula (\ref{eq:Operat-an}) and in the expression for the
transition current the following standard symbols were used:
${\hat{\bf{L}}}=-i[{\bf{r}}\times{\boldsymbol\nabla}]$ is the
operator of the angular momentum, $Y_{LM}({{\bf{n}}})$ are the
spherical functions \cite{Varshalovich-88}, $e_N$, $m_N$ and
$\varphi({\bf{r}})$ are the charge, the mass, and the wave functions
of the nucleon, accordingly, ${\hat{\boldsymbol{\mu}}}$ is the
operator of the magnetic moment of the nucleon:
${\hat{\boldsymbol{\mu}}} = g\mu_N{\hat{\boldsymbol{\sigma}}}$, where
${\hat{\boldsymbol{\sigma}}}$ are the Pauli spin matrices, $\mu_N$ is
the nuclear magneton, $g$ is the empirical magnetic moment of the
nucleon --- $g=2.79$ for the proton and -1.91 for the neutron.

    The spin term of the current ${\bf{j}}_N^s$ in the matrix element
$\langle\hat{\cal{M}}_{1M}^{{m}^{an}}\rangle$ in Eq~(\ref{eq:N-an})
corresponds to the operator in Eq~(\ref{eq:Operat-an}) in square
brackets under the integral ---
$\left[{\bf{r}}({\boldsymbol{\mu}}\cdot{\bf{r}})-
2{\boldsymbol{\mu}}r^2\right]$ (here ${\boldsymbol{\mu}}$ is the
density of the magnetic moment
$\varphi^*_f{\hat{\boldsymbol{\mu}}}\varphi_i$). Let us compare this
operator with a known density operator of the dipole toroid moment
${\bf{t}} = \left[{\bf{r}}({\bf{j}}\cdot{\bf{r}})-
2{\bf{j}}r^2\right]/10$ \cite{Dubovik-90}, that is created by the
convection current ${\bf{j}}$. Operators in square brackets have
different parity and are transformed into each other by exchanging
${\bf{j}}$ and ${\boldsymbol{\mu}}$. Taking into account this fact
let us  introduce the new operator
\begin{equation}
{\bf{c}}_t = \frac{1}{10}
\left[{\bf{r}}({\boldsymbol{\mu}}\cdot{\bf{r}})-
2{\boldsymbol{\mu}}r^2\right]
\label{eq:ct-1}
\end{equation}
and find out what geometrical pattern corresponds to ${\bf{c}}_t$.
Using the spherical functions it is easy to rewrite
Eq~(\ref{eq:ct-1}) in the form:
\begin{equation}
\langle{{\bf{c}}_t}_{M}\rangle = -\frac{\sqrt{\pi}}{3} \int d^3r r^2
\left( {\bf{Y}}^0_{1M}({\bf{n}}) + \frac{\sqrt{2}}{5}
{\bf{Y}}^2_{1M}({\bf{n}}) \right)\cdot{\boldsymbol{\mu}}.
\label{eq:<ct>}
\end{equation}
Then let us remember that the following objects possess the toroid
moment: (a) a toroidal solenoid with a current ${\bf{j}}$; (b) a ring
formed by magnetic moments ${\boldsymbol{\mu}}$ \cite{Dubovik-90}.
For case (b) the following relationship is true:
${\text{curl}}{\bf{t}} ={\boldsymbol{\mu}}$, besides that
$\text{div}{\bf{t}}=0$ due to Helmholtz theorem \cite{Morse-53}.
Substituting ${\text{curl}}{\bf{t}}$ into Eq~(\ref{eq:<ct>}) instead
of ${\boldsymbol{\mu}}$, we get the following relation
$\langle{{\bf{c}}_t}_M\rangle = -i\sqrt{2\pi/3}\int{} d^3r r
{\bf{Y}}^1_{1M}({\bf{n}})\cdot{\bf{t}}$. A sub-integral expression
can be easily transformed to the vector production (see for example,
in Ref.~\cite{Dubovik-90}). As a result, the operator ${\bf{c}}_t$
can be written in as follows: ${\bf{c}}_t =
1/2[{\bf{r}}\times{\bf{t}}]$, in full accordance with similar
expressions for the density of a current magnetic moment
${\boldsymbol{\mu}}_{\bf{j}}=1/2[{\bf{r}}\times{\bf{j}}]$ and for the
toroid moment formed by the ring-like composed magnetic dipoles
${\bf{t}}=1/2[{\bf{r}}\times{\boldsymbol{\mu}}]$ \cite{Dubovik-90}.

\begin{figure}[h]
\includegraphics{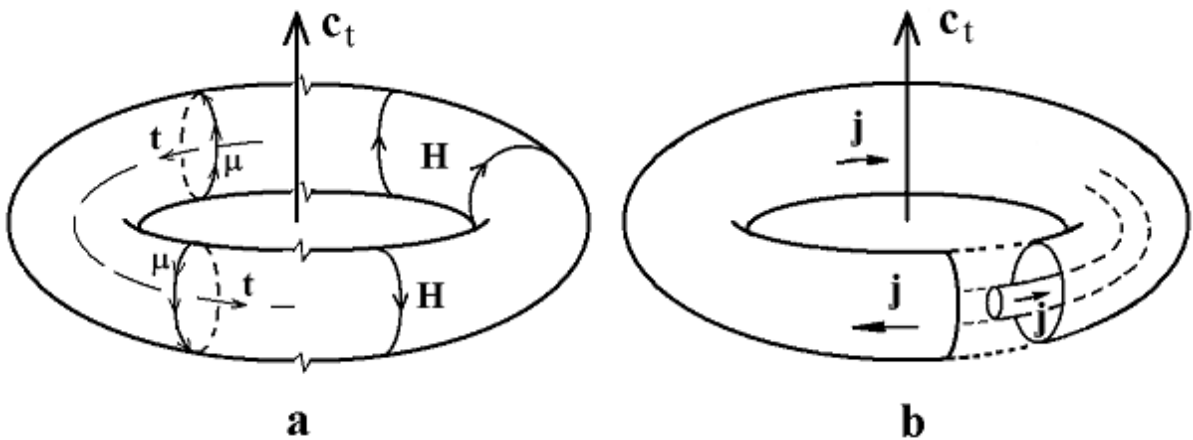}
\caption{Cyclo-toroid.}
\label{Fig1}
\end{figure}

    Expression ${\bf{c}}_t=1/2[{\bf{r}}\times{\bf{t}}]$ results in
$\text{curl}{\bf{c}}_t = {\bf{t}}$. The geometrical image
corresponding to the operator ${\bf{c}}_t$, is the ring-like composed
dipole toroid moments ${\bf{t}}$. Therefore ${\bf{c}}_t$ can be
called the density of cyclo-toroid moment, and
$\langle{\bf{c}}_t\rangle$ --- the cyclo-toroid moment. The
properties of a cyclo-toroid are shown by the torus whose surface is
formed by either ring-like composed magnetic moments
${\boldsymbol{\mu}}$, or by poloidal lines of the magnetic field
(Fig.~\ref{Fig1}a). It is also possible to consider the cyclo-toroid
as a sequence of toroidal solenoids all together forming again a
toroidal solenoid (such objects were considered for the first time in
Ref.~\cite{Afanasiev-96}). In this case the equal currents flow in
opposite directions on the surface of two toruses (one of them is
embedded into another) as it is shown in Fig.~\ref{Fig1}b.

    Zel'dovich's anapole \cite{Zeldovich-57} in the static case
interacts with the external current $j_{ext}$ only. The energy of
this interaction is $-4\pi\int{}d^3r\,{\bf{t}} \cdot{\bf{j}}_{ext}$.
Taking also into account the equality
${\bf{t}}=\text{curl}{\bf{c}}_t$, we arrive straightforwardly to the
Hamiltonian of the interaction of the cyclo-toroid with the external
current:
\begin{equation}
H_{int} = 4\pi\int{}d^3r\,{\bf{c}}_t\cdot\text{curl}{\bf{j}}_{ext}.
\label{eq:Hint}
\end{equation}
That is, the cyclo-toroid interacts locally (only inside the nucleus)
with the vortex part of penetrating external current.

    Let us determine now the role of a cyclo-toroid in the anomalous
conversion. Consider the internal electron conversion from one of the
$s_{1/2}$ states of an atom and show first of all that
$\text{curl}{\bf{j}}\neq0$ during the $M1$ transition of an electron
to the $S_{1/2}$ state of the continuous spectrum.

    We write the electron current in a common way
${\bf{j}}_{fi}=-e{\bar{\psi}}_f{\boldsymbol{\gamma}}\psi_i =
-e{\psi}^*_f{\boldsymbol{\alpha}}\psi_i$, where
${\boldsymbol{\alpha}}=\gamma^0{\boldsymbol{\gamma}}$, $\gamma^0$ and
${\boldsymbol{\gamma}}$ are the Dirac matrixes
\cite{Berestetskii-80}, $e$ is the proton charge, $\psi$ are the
electron wave functions. Let us write these wave functions in the
form of $ \psi({\bf{r}}) = \left( \begin{array}{c}
g(r){\bf\Omega}_{jlm}({\bf{n}}) \\
if(r){\bf\Omega}_{jl'm}({\bf{n}})
\end{array}\right)
$, where $g(r)$ and $f(r)$ are correspondingly the large and the
small components of radial wave functions, ${\bf\Omega}_{jlm}({\bf
n})$ are the spinor spherical harmonics \cite{Berestetskii-80}. For
the bound states in the atom the normalization condition is as
follow: $\int_{0}^{\infty}drr^2(g_i^2(r) + f_i^2(r)) = 1$. The wave
function of the final state is considered as a superposition of the
plane wave and the converging spherical wave.

    For the states with $j=1/2$, $l=0$ and $l'=1$ the angular part
$\psi_{m=\pm1/2}({\bf{r}})$ is very simple, and the components of the
bispinor can be written in the following way: $ g(r)/\sqrt{4\pi}
\left(\begin{array}{c}
\delta_{m1/2}\\
\delta_{m-1/2}
\end{array}\right)
$,
$
if(r)/\sqrt{4\pi} \left(\begin{array}{c}
-\cos\theta\\
-\sin\theta\exp{(i\varphi)}
\end{array}\right)\delta_{m1/2}
$
and
$
if(r)/\sqrt{4\pi} \left(\begin{array}{c}
-\sin\theta\exp{(-i\varphi)}\\
\cos\theta
\end{array}\right)\delta_{m-1/2}
$. Within the spherical coordinate system the current is as follows
\begin{eqnarray}
\displaystyle{}
j_r &=& i\frac{-e}{4\pi}(g_{i}f_{f}-g_{f}f_{i})
\delta_{m_{i}m_{f}},\nonumber\\
j_\theta &=& \mp{i}\frac{-e}{4\pi}e^{\pm{i}\varphi}
(g_{i}f_{f}+g_{f}f_{i})\delta_{m_{i}-m_{f}},\label{eq:j}\\
j_\varphi &=& \frac{-e}{4\pi}(g_{i}f_{f}+g_{f}f_{i})\times
\left\{\begin{array}{l}\displaystyle{}
\mp\sin\theta\delta_{m_{i}m_{f}}\\
\cos\vartheta{} e^{\pm{i}\varphi} \delta_{m_{i}-m_{f}}
\end{array}\right.  \nonumber
\end{eqnarray}
for $m_i=\pm1/2$ correspondingly. We can see that the existence of
the $s_{1/2}\rightarrow{}S_{1/2}$ transition electron current itself
is a totally relativistic effect. The current disappears (as it
should be) when the small parts of the Dirac wave functions $f(r)$
tend to zero.

    The interaction of the current with the cyclo-toroid does not
occur beyond the nucleus range. Therefore, for the radial components
of the  $s_{1/2},S_{1/2}$ states one can use the expressions which
are true for small values of arguments: $g_i(r)=c_i/a_B^{3/2}$,
$f_i(r)=(c_i/a_B^{3/2})(f^{(0)}_i/a_B)r$, and $g_f(r)=c_f/a_B$,
$f_f(r)=(c_f/a_B)(f^{(0)}_f/a_B)r$, where $a_B$ is the Bohr radius,
$c_{i,f}$ and $f^{(0)}_{i,f}$ are the coefficients of decomposition.
(Here the potential inside the nucleus corresponds to the nuclear
charge uniformly distributed over the spherical volume of radius
$r_0$ \cite{Tkalya-94}.) Using the expansion, we can get from
Eq~(\ref{eq:j}) for ${\text{curl}}{\bf{j}}$ the following
\begin{equation}
{\text{curl}}{\bf{j}} = \pm2\frac{-e}{4\pi} \frac{c_i}{a_B^{3/2}}
\frac{c_f}{a_B} \frac{f^{(0)}_i+f^{(0)}_f}{a_B} \times\left\{
\begin{array}{l}
\sqrt{2} {\bf{e}}_{\pm1}\\
-{\bf{e}}_{0}
\end{array}
\right. \,,
\label{eq:curlj}
\end{equation}
where the ${\text{curl}}{\bf{j}}$ vector components are given in the
polar system \cite{Varshalovich-88} and correspond to transitions
$m_i=\pm1/2, m_f=\mp1/2$ and $m_i=m_f=\pm1/2$ accordingly. Since
$g(0)\neq0$ and $df(r)/dr|_{r\rightarrow0}\neq0$ for $s_{1/2},
S_{1/2}$ states, then ${\text{curl}}{\bf{j}}\neq0$.

    As an example we consider the known anomaly in the conversion
$M1$ transition with the energy of $\omega_N=482$ keV in $^{181}$Ta
nucleus. The experimental magnitude of the $K$ shell conversion
coefficient is 3 times as great as the so-called normal theoretical
magnitude \cite{Firestone-91}. The magnitude of the penetration
parameter $\lambda^{(0)}= \langle \hat{\cal{M}}_{1M}^{{m}^{an}}
\rangle / \langle \hat{\cal{M}}_{1M}^{m}\rangle$ equals to 150
\cite{Firestone-91}. The transition occurs between the states with
quantum numbers in the Nilsson model being as follows:
$J^{\pi}K[Nn_z\Lambda]\Sigma = 5/2^+5/2[402]\uparrow$ and
$7/2^+7/2[404]\downarrow$ \cite{Nilsson-58}. The radiation nuclear
matrix element of this transition is forbidden by asymptotic quantum
numbers, while the anomalous one is allowed \cite{Nilsson-58}.
However, the selection rules for the operator $r^2{\hat{\bf{L}}}$ in
Eq~(\ref{eq:Operat-an}) forbid the anomaly nuclear transition in
$^{181}$Ta \cite{Nilsson-58}. Therefore we can infer that the anomaly
results just from the interaction of the transition cyclo-toroid
moment with the vortex part of the penetrating electron current.

    In order to determine $\langle{\bf{c}}_t\rangle$ let us use the
standard view for the radiation nuclear matrix element
\cite{Eisenberg-70}: $\langle \hat{\cal{M}}_{LM}^{m}\rangle =
i\omega^L/(2L+1)!! \sqrt{(L+1)/L} C^{J_fM_f}_{J_iM_iLM}
\langle{}J_f\|\hat{\cal{M}}_{L}^{m}\|J_i\rangle/\sqrt{2J_f+1}$, where
$C^{J_fM_f}_{J_iM_iLM}$ are the Klebsch-Gordan coefficients. The
reduced matrix element
$\langle{}J_f\|\hat{\cal{M}}_{L}^{e/m}\|J_i\rangle$ in the nuclear
spectroscopy is usually related to the experimentally measured
reduced probability of the nuclear transition
$B(E/ML;J_i\rightarrow{}J_F) =
|\langle{}J_f\|\hat{\cal{M}}_{L}^{e/m}\|J_i\rangle|^2/(2J_i+1)$. The
reduced probability of the $M1$ transition in the single-particle
Weisskopf model is $B(M1;W)=45\mu_N^2/8\pi$ \cite{Eisenberg-70}. As a
rule, the value of the experimental probability is given in Weisskopf
units $B_{W.u.}=B(E/ML;J_i\rightarrow{}J_F)/B(E/ML;W)$. For the
considered transition in $^{181}$Ta it is
$B_{W.u.}=6.21\times10^{-7}$ \cite{Firestone-91}.

    Let us substitute operator in square brackets in
Eq~(\ref{eq:Operat-an}) to operator $10{\bf{c}}_t$ and then, using
the definitions for $\lambda^{(0)}$ and $\langle
\hat{\cal{M}}_{LM}^{m}\rangle$, derive the expression for the reduced
matrix element of the cyclo-toroid moment of the transition through
the parameters measured in the experiment:
\begin{equation}
\langle \|{{\bf{c}}_t}\|\rangle =
c_t^{(0)}\lambda^{(0)}\,\sqrt{\frac{6}{5}(2J_i+1)B_{W.u.}}.
\label{eq:Ct-reduced}
\end{equation}
We introduced here $c_t^{(0)} = \mu_N r_0^2/4$ as the ``unit'' for
measurement of the cyclo-toroid moment. For $^{181}$Ta nucleus the
following is true: $c_t^{(0)} =0.52$ fm$^2$/GeV if the radius of a
nucleus with atomic number $A$ is $r_0= 1.2A^{1/3}$ fm. Using
Eq~(\ref{eq:Ct-reduced}) one can estimate the value
$\langle\|{\bf{c}}_t\|\rangle$ for the anomalous conversion $M1$(482
keV) transition: $\langle\|{\bf{c}}_t\|\rangle\simeq0.3\,c_t^{(0)}$.

\section{3. Static nuclear cyclo-toroid moment}

    The cyclo-toroid moment is a pseudovector. Atomic nuclei, like
some other objects, for example, nucleons, can have a static
cyclo-toroid moment in the ground state. The value of a ${\bf{c}}_t$
for nucleus should not depend on the Fermi coupling constant $G_F$ of
the weak interaction, unlike the static toroid moment (the latter as
a rule arises due to the parity violation for the nuclear forces and
contains $G_F$ \cite{Zeldovich-57,Flambaum-80}). Another positive
aspect is the geometric factor similar to the one considered for
toroid moments in Ref.~\cite{Flambaum-80}. The cyclo-toroid moment
$\langle{\bf{c}}_t\rangle$ is proportional to the nucleus volume,
that is, to the atomic number $A$. As for the interaction of the
cyclo-toroid moment with an atom electron, this interaction is of the
electromagnetic nature and, therefore, is proportional to the fine
structure constant $e^2$.  Besides, the additional smallness of
$\sim{}e^2$ is produced by the small component of the Dirac wave
function $f(r)$.

    The interaction of the atomic $s$-electron with the cyclo-toroid
moment of a nucleus results in the contribution to the atomic level
energy already in the first order of the perturbation theory. To
demonstrate this, let us estimate the shift of the $6s$ level in
$^{133}$Cs. The dependence of the energy on the electron spin
projection $\pm1/2$ and on the value of the cyclo-toroid moment $c_t$
(here and further $c_t$ is measured in $c_t^{(0)}$ units) can be
found from the following formula:
\begin{equation}
\Delta{}E_{ns} = \pm{}c_{t}m_{e} e^6 \frac{m_e}{2m_N}
\left(\frac{r_0}{a_B}\right)^2 c^2_{ns}f^{(0)}_{ns},
\label{eq:DeltaE}
\end{equation}
that can be obtained from Eqs~(\ref{eq:Hint}) and (\ref{eq:curlj}),
if one uses the $ns_{1/2}$ wave function for the initial and the
final state simultaneously, and takes into account the relation
$a_B=(m_{e}e^2)^{-1}$, where $m_e$ is the electron mass. We have
$|\Delta{}E_{6s}| = 0.7\times10^{-7}c_t$ eV when $c_{6s} = 8.7$,
$f^{(0)}_{6s} = -1.2\times10^{3}$.

    Hydrogen-like ions in the $1s_{1/2}$ state are perspective objects
for such investigation. We give here the qualitative estimations for
the H-like ions of two heavy nuclei:
$|\Delta{}E_{1s}(^{133}$Cs$^{54+})|\sim 1\times10^{-3}c_t$ eV and
$|\Delta{}E_{1s}(^{233,235}$U$^{91+})\sim 1\times10^{-1}c_t$ eV.

    The proton constituted of quarks can also possess a cyclo-toroid
moment. In this case the shift of $1s$ level of a hydrogen atom as a
function of $c_t$ is $|\Delta{}E_{1s}| \simeq 1\times10^{-11}c_t$ eV
(the calculation was performed according to formula (\ref{eq:DeltaE})
with the following parameters: $c_{1s}=2$,
$f^{(0)}_{1s}=-Z/(2m_{e}r_0)=-223.6$ if $r_0=0.862$ fm, $Z=1$). It
should be noted that the magnitude of $\Delta{}E_{1s}$ for
$c_t\simeq1$ is at present close to the accuracy of the measurement
of the Lamb shift for the $1s$ level of hydrogen \cite{Berkeland-95}.

    In conclusion it is important to note the following. The system
depicted in Fig.~\ref{Fig1} has no charges. Therefore, for all the
charge moments there is
$Q^q_{LM}\sim\int{}d^3rr^LY^*_{LM}({\bf{n}})\rho(r)=0$ (where $\rho$
is the charge density). For all the magnetic moments of the system
there is also
$Q^m_{LM}\sim\int{}d^3rr^LY^*_{LM}({\bf{n}}){\text{div}}
{\boldsymbol{\mu}}=0$, since all the elementary magnetic moments are
ring-like composed and can be written in the following way:
${\boldsymbol{\mu}}= {\text{curl}}{\bf{t}}$, and
${\text{div}}\,{\text{curl}}{\bf{t}}=0$. The same is also true for
the toroid moment $Q^t_{LM}\sim
\int{}d^3rr^LY^*_{LM}({\bf{n}}){\text{div}}{\bf{t}}$
\cite{Dubovik-90}, because for the object in Fig.~\ref{Fig1} there is
${\bf{t}}={\text{curl}}{\bf{c}}_t$. Thus, the system has no toroid
moments. In spite of the fact that the mentioned electromagnetic
moments are equal to zero, the interaction energy of the system with
the external current has a nonzero value. So, the cyclo-toroid moment
can be the single non-vanishing moment of the quantum object.

\section{Acknowledgement}

    During recent years the author discussed some problems
considered in Sec.~2 of this paper with Dr.
M.A.~Listengarten\footnote{Deceased January 2004.}.  Without his
benevolent interest, constructive advices and persistent wish to see
work finished, this paper would be hardly written. This work was
partially supported by the ISTC, Project N2651, and by the Leading
Science School Grant N2078.2003.2.

\end{document}